\documentclass{moriond}

\newcommand{\ep}{\epsilon}

\newcommand{\al}{\alpha}
\newcommand{\bt}{\beta}
\newcommand{\g}{\gamma}

\newcommand{\Or}{\mathcal O}

\newcommand{\vL}{\ensuremath{\mathcal{L}}}
\newcommand{\vp}{\varphi}
\newcommand{\sq}{^{2}}

\newcommand{\dslash}[1]{#1 \llap{/\kern-0.5pt}}
\newcommand{\Dslash}[1]{#1 \llap{/\kern+1.5pt}}
\newcommand{\DDslash}[1]{#1 \llap{/\kern+2.3pt}}
\newcommand{\dslashh}[1]{#1 \llap{/\kern+1pt}}

\newcommand{\Ex}[1]{\cdot 10^{#1}}
\newcommand{\bea}{\begin{eqnarray}}
\newcommand{\eea}{\end{eqnarray}}
\newcommand{\bma}{\begin{pmatrix}}
\newcommand{\ema}{\end{pmatrix}}
\newcommand{\nn}{\nonumber}

\newcommand{\Photo}{}

\begin{document}
\vspace*{2cm}
\title{$\ep'$ from right-handed currents}

\author{ W. Dekens }

\address{
Theoretical division, Los Alamos National Laboratory, Los Alamos, NM 87545, USA\\
New Mexico Consortium, Los Alamos Research Park, Los Alamos, NM 87544, USA
}

\maketitle\abstracts{
Recent lattice determinations of direct CP violation in kaon decays, $\ep'$, suggest a discrepancy of several sigma between experiment and the standard model.
Assuming that this situation is due to new physics, we investigate a solution in terms of  right-handed charged currents. Chiral perturbation theory, in combination with lattice QCD results, allows one to accurately determine the effect of right-handed interactions on $\ep'$. In addition, similar techniques provide a direct link between the right-handed contributions to $\ep'$ and hadronic electric dipole moments.
We demonstrate that the $\ep'$ discrepancy can be resolved with right-handed charged currents, and that this scenario can be falsified by next-generation hadronic electric dipole moment experiments
}

\section{Introduction}
Although the direct CP violation in $K_L\to \pi\pi$ decays, $\ep'$, has been precisely measured over a decade ago~\cite{AlaviHarati:2002ye,Abouzaid:2010ny,Batley:2002gn}, the corresponding standard model (SM) predictions have not yet reached a comparable precision. Such a prediction is a challenging task that 
requires the calculation of nonperturbative  matrix elements. 
Recently, these matrix elements have been determined using lattice QCD~\cite{Bai:2015nea}, and suggest a $2-3\, \sigma$ discrepancy between the SM and the experimental value. 
This discrepancy is in agreement with the results of Refs.\ \cite{Buras:2015xba,Buras:2016fys,Kitahara:2016nld}, while several analytic approaches find values for $\ep'$ that are  consistent with experiment~\cite{Bertolini:2000dy,Bijnens:2000im,Pallante:2001he}. 

Assuming that this tension survives future improved lattice determinations, it is interesting to investigate possible explanations in terms of new physics. 
Several explanations in terms of vector-like quarks~\cite{Bobeth:2016llm}, $331$ models~\cite{Buras:2015kwd}, $Z^{(\prime)}$ couplings~\cite{Buras:2015jaq}, and supersymmetric scenarios~\cite{Kitahara:2016otd,Endo:2016aws,Crivellin:2017gks} have been discussed in the literature.
 Here we investigate a scenario involving a single gauge-invariant dimension-six operator~\cite{Cirigliano:2016yhc}. This  operator induces right-handed charged currents (RHCCs), which couple the $W$ boson to right-handed quarks,
 and is given by
\bea \label{eq:xiOperator}
\vL_{\rm eff}  = {\cal L}_{\rm SM} +  \frac{2}{v^2} i \tilde{\vp}^{\dagger} D_{\mu} \vp \, \bar{u}^i_R \gamma^\mu \,\xi_{i j} d^j_R 
+  \mathrm{h.c.}, 
  \ \to \  { \cal L}_{\rm SM} + \frac{g}{\sqrt 2}\bigg[\xi_{ij}\,\bar u^i_R \g^\mu  d^j_R\,W_\mu^+   \bigg]\left(1+\frac{h}{v}\right)^2 +{\rm h.c.},
\ \ \ 
\eea
where $\vp$ is the Higgs doublet, $v\approx 246$ GeV is its vacuum expectation value (vev), and the covariant derivative is given by $D_\mu =\partial_\mu -i g/2 \tau\cdot W_\mu -ig'/2 B_\mu$, with $g$ and $g'$ the $SU(2)$ and $U(1)_Y$ gauge couplings. Finally, $\xi_{ij}$ is a $3\times 3$ matrix in flavor space, whose elements are expected to scale as $v\sq/\Lambda\sq$, where $\Lambda$ is the scale of new physics.
This interaction is generated in left-right symmetric models~\cite{Mohapatra:1974hk,Senjanovic:1975rk}, 
which are based on the gauge group $SU(2)_L\times SU(2)_R\times U(1)_{B-L}$.  These models feature a new right-handed $W_R$ boson which can undergo mass-mixing with  the SM $W$-boson. After integrating out the heavy $W_R$ this mixing induces Eq.\ \ref{eq:xiOperator}. Explicitly one has, $\xi_{ij}\approx \frac{g_R\sq}{2}\frac{\kappa\kappa'}{m_R\sq}e^{i\al}(V_R)_{ij}$, where $\kappa,\, \kappa'\sim v$ are the magnitudes of the vevs that break $SU(2)_L$ and $\al$ is the phase difference between them, $m_R$ and $g_R$ are mass of the $W_R$ boson and its gauge coupling, while $V_R$ is the right-handed analogue of the CKM matrix.

Here we do not restrict to a specific model and focus on two elements of the $\xi$ matrix~\footnote{For the phenomenology of the remaining flavor structures see~\cite{Alioli:2017ces}.}, namely $\xi_{ud}$ and $\xi_{us}$. After integrating out the $W$ boson these elements induce both $\Delta S=1$ and $\Delta S=0$ four-quark operators, which contribute to $\ep'$ and hadronic EDMs, respectively. Both contributions depend on nonperturbative matrix elements. 
In the case of $\ep'$ chiral symmetry allows one to relate the necessary matrix elements to those of SM operators calculated on the lattice. 
Although the EDM analysis depends on several nonperturbative quantities, not all of which are known, we argue that the same matrix elements allow one to estimate the leading contributions in this case as well.

We discuss the low-energy Lagrangian induced by Eq.\ \ref{eq:xiOperator} in section \ref{sec:LowELag}. The impact of this Lagrangian on $\ep'$ and hadronic EDMs are derived in section \ref{sec:eps} and \ref{sec:EDMs}, respectively. We discuss the resulting constraints  and the possibility of a solution to the $\ep'$ discrepancy in section \ref{sec:disc}.

\section{Low-energy Lagrangian}\label{sec:LowELag}
\subsection{Quark-level Lagrangian}
After integrating out the heavy SM fields the couplings $\xi_{ud}$ and $\xi_{us}$ of Eq.\ \ref{eq:xiOperator} give rise to the following four-quark interactions
\begin{eqnarray}\label{eq:4q4}
\mathcal L_{LR} &=& -  \sum_{i=1}^2\Big(   \, C^{ud\, ud}_{i\, LR} \, \mathcal O^{ud\, ud}_{i\, LR}  + \, C^{us\, us}_{i\, LR}  \,  \mathcal O^{us\, us}_{i\, LR}+ C^{us\, ud}_{i\, LR} \, \mathcal O^{us\, ud}_{i\, LR}   + \, C^{ud\, us}_{i\, LR}  \,  \mathcal O^{ud\, us}_{i\, LR}  +{\rm h.c.}\Big) \ ,
\end{eqnarray}
where,
 $\mathcal O^{ij\, lm}_{1\, LR} = \bar d^m \gamma^\mu P_L u^l \, \bar u^i \gamma_\mu P_R d^j$  and $\mathcal O^{ij\, lm}_{2\, LR}  =\bar d_\al^m   \gamma^\mu P_L u_\bt^l \, \bar u_\bt^i  \gamma_\mu P_R d_\al^j
$, with
$\al,\,\bt$  color indices.  These `left-right' operators violate CP as long as the corresponding couplings have an imaginary part. The first two $\Delta S=0$ operators in Eq.\ \ref{eq:4q4} will induce EDMs, while the second pair violate strangeness by one unit and contribute to $\ep'$. Note that no $\Delta F=2$ operators are generated at tree level.
The matching at the $W$ boson mass scale gives,
\begin{eqnarray}\label{eq:4q2}
C^{ij\, lm}_{1\, LR}(m_W)  =\frac{4 G_F}{\sqrt{2}}  V^*_{lm} \xi_{i j}\ , \qquad C^{ij\, lm}_{2\, LR}(m_W) = 0 \ ,
\end{eqnarray}
while the $\mathcal O_{2\, LR}$ operators are induced through QCD renormalization.

After evolving the above Lagrangian to $\mu\approx 3$ GeV, the contributions to $\ep'$ and EDMs still require the matrix elements of the left-right operators, which we obtain from ChiPT. Before moving on to the chiral realization of the left-right operators, however, we slightly rewrite the relevant parts of the effective Lagrangian,
\begin{equation}\label{eq:LagGev}
\mathcal L = \mathcal L^{\rm QCD}_{m_q = 0} - \bar q \mathcal M q + \bar q \left[ m_3 t_3 + m_6 t_6 + m_8 t_8\right]i \gamma_5  q + {\cal L}_{LR}\ ,
\end{equation}
where $t^a$ are the $SU(3)$ generators, $q$ is a triplet of quark fields $q = (u, d, s)$, and $\mathcal M = {\rm diag}(m_u,\,m_d,\, m_s)$. In addition, we assumed that the strong CP problem is solved by a Peccei-Quinn mechanism~\cite{Peccei:1977hh}. Finally, the left-right operators induce couplings which couple the neutral mesons,  $\pi^0, K^0, \eta$, to the vacuum. To avoid such couplings  when constructing the Chiral Lagrangian we perform a $SU(3)_L\times SU(3)_R$ rotation to eliminate these terms, this introduces $m_{3,6,8}$ which are specified in the next section.

\subsection{Chiral Lagrangian}
To construct the chiral Lagrangian it is useful to note that the left-right operators can  schematically be written as, $(\bar q \g_\mu t^a P_Lq)\, (\bar q \g^\mu t^b P_Rq)$. Such operators belong to the $8_L\times 8_R$ representation of $SU(3)_L\times SU(3)_R$ and so transform as $t^a\to L t^a L^\dagger,\, t^b\to R t^b R^\dagger$, where $L,\, R \in SU(3)_{L,R}$. Using these transformation properties and the well-known chiral realization of the usual QCD Lagrangian, the leading-order mesonic Lagrangian is given by
\begin{eqnarray}\label{ChiLag}
& & \mathcal L_\pi = \frac{F_0^2}{4} \textrm{Tr} \left(\partial_\mu U \partial^\mu U^\dagger  \right) +\frac{F_0^2}{4} \textrm{Tr} \left( U \chi^\dagger + U^\dagger \chi \right) 
\nn\\
&& + \frac{F_0^4}{4}   \textrm{Tr}\left( U^\dagger t^b Ut^a  \right)\sum_{i=1,2} \mathcal A_{i\, LR}\bigg[
C_{i\, LR}^{ud\, ud} (\delta_{a1}-i\delta_{a2})(\delta_{b1}+i\delta_{b2})+C_{i\, LR}^{us\, us}(\delta_{a4}-i\delta_{a5})(\delta_{b4}+i\delta_{b5})\nn\\
&&+C_{i\, LR}^{ud\, us}(\delta_{a4}-i\delta_{a5})(\delta_{b1}+i\delta_{b2})+C_{i\, LR}^{us\, ud}(\delta_{a1}-i\delta_{a2})(\delta_{b4}+i\delta_{b5})+{\rm h.c.}\bigg]\ ,
\end{eqnarray}
here $F_0$ is the pion decay constant in the Chiral limit, and $U$ is the usual matrix of pseudo Nambu Goldstone bosons in the notation of~\cite{Cirigliano:2016yhc}.
Furthermore,
\begin{equation}\label{chi}
\chi = 2 B \left( \mathcal M + i \left(   m_3  t_3  +m_6t_6+ m_8 t_8  \right)\right)\, ,
\end{equation}
with
\begin{eqnarray}\label{align}
m_3 &=& - \sum_{i=1,2}r_i {\rm Im}\,\left( C_{i\, LR}^{ud\, ud} + \frac{1}{2} C_{i\, LR}^{us\, us} \right) ,\qquad m_6 =  \frac{1}{2}\sum_{i=1,2}r_{i}{\rm Im}\,\left(C_{i\, LR}^{ud\, us}+ C_{i\, LR}^{us\, ud}\right), \nn\\
m_8 &=& -\frac{\sqrt{3}}{2}\sum_{i=1,2}  r_i \textrm{Im} \, C_{i\, LR}^{us\, us}\ ,
\end{eqnarray}
where $r_{i}=\frac{F_0\sq}{B} \mathcal A_{i\, LR}$.

The second and third lines in the Lagrangian in Eq.\ \ref{ChiLag} involve terms that will induce EDMs and $\ep'$, respectively. These contributions depend on the low energy constants (LECs), $\mathcal A_{i\, LR}$, which can be related to the known matrix elements of the SM electroweak penguin operators, $\mathcal Q_7$ and $\mathcal Q_8$~\cite{Bijnens:1983ye,Buchalla:1995vs}. 
These SM operators 
also transform as $8_L\times 8_R$ and therefore induce similar terms as $\mathcal O_{i\, LR}$ in the Chiral Lagrangian, with the same LECs, $\mathcal A_{i\, LR}$. As a result, we can express their matrix elements for $K_L\to \pi\pi$ in terms of $\mathcal A_{i\, LR}$. At leading order (LO) in ChiPT, together with recent lattice results~\cite{Blum:2012uk}, this gives
\begin{eqnarray}
\mathcal A_{1\, LR}(3\, \textrm{GeV}) &=& \frac{1}{\sqrt{3} F_0}  \, \langle (\pi\pi)_{I=2}|\mathcal  Q_7|K^0\rangle
 + \mathcal O\left(m^2_K\right) \simeq (2.2\pm0.13) \, {\rm GeV}^2\ ,\nn\\
\mathcal A_{2\, LR}(3\, \textrm{GeV}) &=& \frac{1}{\sqrt{3} F_0}  \, \langle (\pi\pi)_{I=2}|\mathcal  Q_8|K^0\rangle
 + \mathcal O\left(m^2_K\right)\simeq (10.1\pm0.6) \, {\rm GeV}^2\ . 
\label{LEC}
\end{eqnarray}

\section{Contribution to CP violation in the kaon sector}\label{sec:eps}
\subsection{Direct CP violation}
The measure of CP direct violation in $K_L\to \pi\pi$ decays is given by,
\bea
{\rm Re}\, \bigg(\frac{\ep'}{\ep}\bigg) = {\rm Re}\,\bigg(\frac{i \omega e^{i(\delta_2-\delta_0)}}{\sqrt{2}\ep}\bigg)\bigg[\frac{{\rm Im}\,A_2}{{\rm Re}\,A_2}-\frac{{\rm Im}\,A_0}{{\rm Re}\,A_0}\bigg]\label{epsPrime}\ , 
\eea
where $A_{0,2} e^{i\delta_{0,2}}$ are the amplitudes for final states with total isospin $I=0,2$,  $\omega = {\rm Re}\,A_2/{\rm Re}\,A_0$, and $\ep$ denotes the CP violation in $\bar K-K$ mixing.
The left-right operators  contribute to the imaginary parts of the amplitudes, ${\rm Im}\,A_{0,2}$, while the remaining  quantities in the above expression are well known experimentally. The Chiral Lagrangian of the previous section, Eq.\ \ref{ChiLag}, together with the determination of the LECs, Eq.\ \ref{LEC}, now allows us to calculate the contributions to these amplitudes. One would expect such a calculation to be subject to $\Or(m_K\sq/\Lambda_\chi\sq)$ corrections due to the fact that Eq.\ \ref{LEC}  is  a LO ChiPT prediction.
Luckily, this is not the case for the $I=2$ amplitude. The reason is that, after an isospin decomposition, the $I=3/2$ parts of the $O_{1(2)\, LR}$ and $\mathcal Q_{7(8)}$ coincide. Thus, the right-handed contributions to the $I=2$ amplitudes can be determined up to isospin corrections, which gives
\bea
{\rm Im}\, A_2(\xi) &=& \frac{1}{6\sqrt{2}}{\rm Im }\, \bigg[\left(C_{1\, LR}^{udus}-C_{1\, LR}^{usud\,*}\right)\langle (\pi\pi)_{I=2}|\mathcal Q_7|K^0\rangle\nn\\
&&+\left(C_{2\, LR}^{udus}-C_{2\, LR}^{usud\,*}\right)\langle (\pi\pi)_{I=2}|\mathcal Q_8|K^0\rangle\bigg].
\eea
Unfortunately, the $I=1/2$ parts of the left-right operators do not coincide with those of the SM operators, but at LO in ChiPT we find, ${\rm Im}\, A_0(\xi) = -2\sqrt{2}{\rm Im}\,A_2(\xi)$.
In total we then use Eq.\ \ref{epsPrime} with ${\rm Im}\, A_{0,2}= {\rm Im}\,A_{0,2}^{\rm SM}+ {\rm Im}\,A_{0,2}(\xi)$ and employ 
the results of~\cite{Bai:2015nea,Buras:2015yba} for the SM prediction. It should be noted that although 
$A_0(\xi)$ is only known up to $\Or(m_K\sq/\Lambda_\chi\sq)$ corrections it gives a subleading contribution to $\ep'$, as it is suppressed by the $\Delta I=1/2$ rule. We expect the main source of uncertainties to result from the lattice determinations of the matrix elements.

\subsection{CP violation in mixing}
Apart from the direct CP violation in kaon decays, the right-handed interactions can also induce CP violation in mixing, $\ep_K$. Although $\xi_{ud}$ and $\xi_{us}$ do not induce tree-level $\Delta S=2$ operators, they do contribute to $\ep_K$ through  short- and long-distance effects. The former arise through  box diagrams involving the $\xi$ couplings. However, due to the chirality of the vertices, the box diagrams linear in $\xi$ require one internal and one external quark mass insertion, i.e.\ they are suppressed by $\frac{m_u m_s}{m_W\sq}$. The short-distance contributions are therefore negligible.

The long-distance contribution arises from the combination of a $\Delta S=1$ left-right interaction with a $\Delta S=1$ SM charged current. The Chiral realizations of these operators lead to diagrams where $K^0$ mixes into a pion or eta meson, which then mixes into a $\bar K^0$. 
This involves the LECs  of the left-right operators, $\mathcal A_{i\, LR}$, as well as those for the weak charged current~\cite{Cirigliano:2011ny}. Here we follow \cite{Cirigliano:2016yhc} and estimate this contribution to $\ep_K$ by the tree-level diagrams (which are non-zero at NLO) and assign a $50\%$ uncertainty to it due to  unknown NLO counterterms.
\section{Contribution to hadronic EDMs}\label{sec:EDMs}
The contributions of the left-right operators to hadronic EDMs can be calculated by first matching to an extension of  chiral effective field theory that contains CP-violating hadronic interactions \cite{deVries:2012ab,Bsaisou:2014oka}.
Chiral power counting then predicts~\cite{deVries:2012ab,Bsaisou:2014oka} that contributions of the four-quark operators to nuclear EDMs are dominated by  long-range pion-exchange between nucleons \footnote{Note that chiral power counting has not been tested for systems as large as $^{199}$Hg or $^{225}$Ra.}. 
The leading  pion-nucleon couplings, $\bar g_{0,1}$, are induced by the left-right operators in several ways. Firstly, there is a direct contribution whose LEC involves matrix elements of the form, $\langle N\pi |O_{i\, LR}|N\rangle$, which are currently unknown. A second contribution arises due to the rotation performed to align the  vacuum, mentioned in section \ref{sec:LowELag}. 
The relevant meson-baryon Lagrangian then takes the following form,
\begin{eqnarray}\label{eq:L2}
\mathcal L_{\pi N} &=&   b_0 \textrm{Tr} \left(\bar B_{} B\right) \textrm{Tr} \chi_+  +  b_D \textrm{Tr}\left(\bar B \{ \chi_+, B \}  \right) + b_F \textrm{Tr}\left(\bar B [ \chi_+, B ]  \right) 
+ \mathcal L_{\rm direct} \, ,
\end{eqnarray}
where $B$ represents the octet of baryon fields, notation is as in~\cite{Cirigliano:2016yhc}, and $\chi_+ = u^{\dagger} \chi u^{\dagger} + u \chi^{\dagger} u$.

Here the direct contributions to $\bar g_{0,1}$, with unknown LECs, are due to $\vL_{\rm direct}$. The  contributions induced by vacuum alignment  arise through $\chi_+$ and depend on $\mathcal A_{i\, LR}$, and $b_{0,D,F}$. Since $b_{0,D,F}$ can be related to the baryon mass splittings~\cite{Aoki:2016frl,Borsanyi:2013lga,Borsanyi:2014jba,Brantley:2016our}, and $\mathcal A_{i\, LR}$ are known from Eq.\ \ref{LEC} this contribution can be estimated reliably. Using the conventions of~\cite{Cirigliano:2016yhc} for $\bar g_{0,1}$, the indirect contributions, including lattice uncertainties, give
\begin{eqnarray}\label{couplings0}
\frac{\bar{g}_0}{2 F_\pi} &=& - (0.16 \pm 0.03)  \times 10^{-5}\, \textrm{Im} (V^*_{us} \xi_{us}) \, ,\nn\\
\frac{\bar{g}_1}{2 F_\pi} &=& - \left( 2.9 \pm 0.33 \right)\times 10^{-5} \,  \textrm{Im} (V^*_{us} \xi_{us})   -  \left(  5.7 \pm 0.67    \right) \times 10^{-5}\,  \textrm{Im} (V^*_{ud} \xi_{ud})\, .
\end{eqnarray}
Naive-dimensional-analysis estimates for the additional direct pieces suggest they are roughly an order of magnitude smaller than the indirect contributions. As such, we follow \cite{Cirigliano:2016yhc} and use the indirect piece as the central values for $\bar g_{0,1}$ and conservatively assign a $50\%$ uncertainty due to the direct piece. 
\subsection{The neutron EDM}
Although $\bar g_{0,1}$ give the dominant contributions to nuclear EDMs, for the neutron EDM additional counterterms appear at the same order. 
One has \cite{Mereghetti:2010kp,Seng:2014pba}
\begin{eqnarray}
d_n &=& \bar d_n  (\mu) + \frac{e g_A \bar g_1}{(4\pi F_\pi)^2} \left(  \frac{\bar g_0}{\bar g_1} \left( \log \frac{m^2_\pi}{\mu^2} - \frac{\pi m_\pi}{2 m_N} \right)  + \frac{1}{4 } \left( \kappa_1 - \kappa_0\right) \frac{m^2_\pi}{m_N^2} \log \frac{m^2_\pi}{\mu^2}  \right) \ ,
\label{eq:dn}
\end{eqnarray}
where $g_A \simeq 1.27$ is the nucleon axial charge, 
and $\kappa_1 = 3.7$ and $\kappa_0 = -0.12$ are related to the nucleon magnetic moments. 
$\bar d_{n} (\mu) $  is a counterterm, which is again unknown. We estimate its size by the $\mu$ dependence of the loop contributions, which we obtain by varying $\mu$ from $m_K$ to $m_N$ in Eq.\ \ref{eq:dn}. The resulting sizes are in agreement with naive-dimensional-analysis estimates. As a result, we take Eq.\ \ref{eq:dn} as the central value with $\bar d_{n}(\mu)=0$. We estimate the uncertainties by the combination of the errors on $\bar g_{0,1}$ discussed above together with the variation due to the $\mu$ dependence~\cite{Cirigliano:2016yhc}.

\subsection{Nuclear EDMs}
As already mentioned, for nuclear EDMs, the dominant contributions should be captured by $\bar g_{0,1}$, implying that no further unknown LECs enter the expressions in this case.
Nuclear calculations,   within large uncertainties,  predict 
\cite{deJesus:2005nb,Bsaisou:2014zwa,Ban:2010ea,Dzuba:2009kn,Engel:2013lsa,deVries2011b,Bsaisou:2014oka,Singh:2015aba,Yamanaka:2017mef}
\bea\label{eq:NuclEDM}
d_D &=&- (0.18 \pm 0.02) \frac{\bar g_1}{2F_\pi}\, e \, {\rm fm}\ ,\nn\\ 
d_{\mathrm{Hg}} &=& (2.8\pm 0.6)\Ex{-4}\cdot \bigg(0.13^{+0.5}_{-0.07}\, \frac{\bar g_0}{2F_\pi} + 0.25^{+0.89}_{-0.63}\,\frac{\bar g_1}{2F_\pi}\bigg)e\, {\rm fm}\ ,  \nn \\
d_{\mathrm{Ra}} &=& (7.7\pm 0.8)\Ex{-4}\cdot\left(-19^{+6.4}_{-57}\,\frac{\bar g_0}{2F_\pi} + 76^{+227}_{-25}\,\frac{\bar g_1}{2F_\pi}\right)e\, {\rm fm}\ ~.
\eea
We set constraints using Eqs.\ \ref{eq:NuclEDM} and \ref{eq:dn} together with the experimental measurements~\cite{Afach:2015sja,Graner:2016ses,Bishof:2016uqx,Parker:2015yka}.

\section{Discussion}\label{sec:disc}
\begin{figure}[t]
\center
\includegraphics[width=7.5cm]{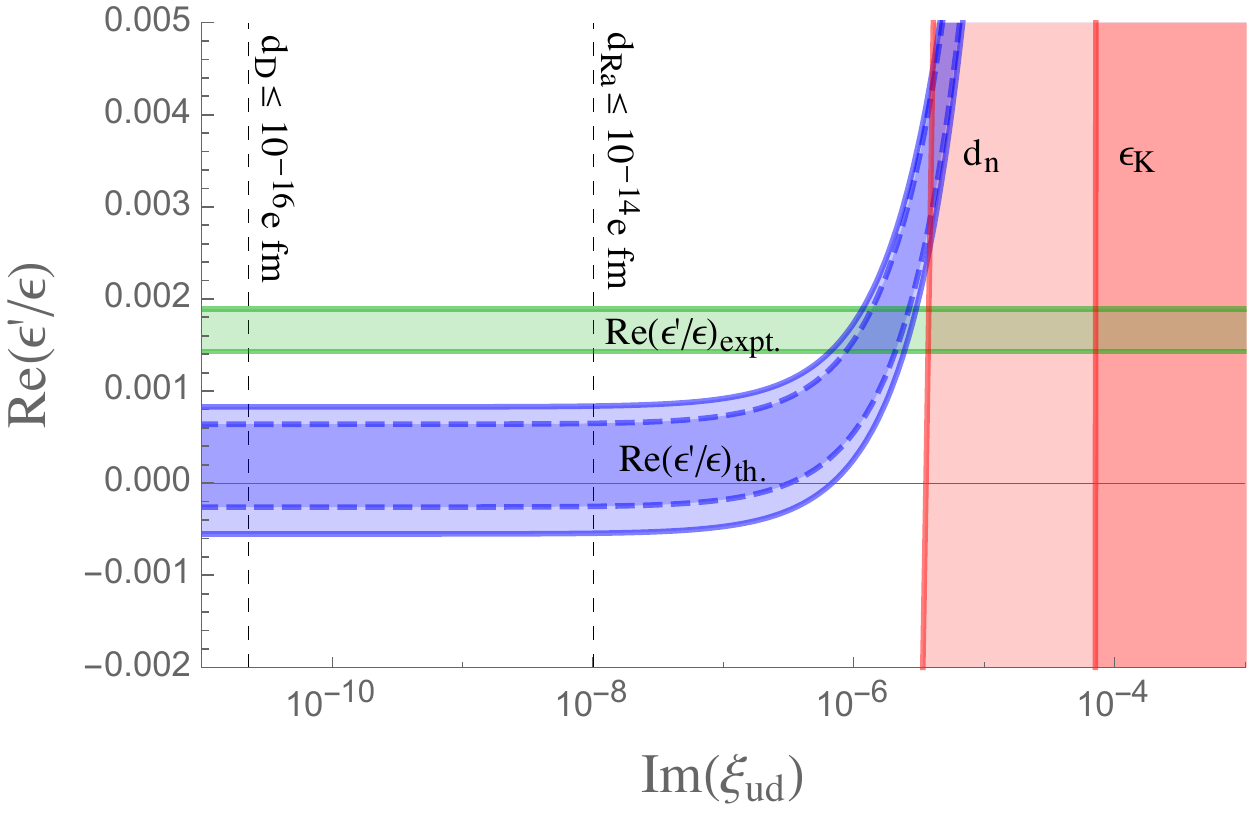}
\quad
\includegraphics[width=7.5cm]{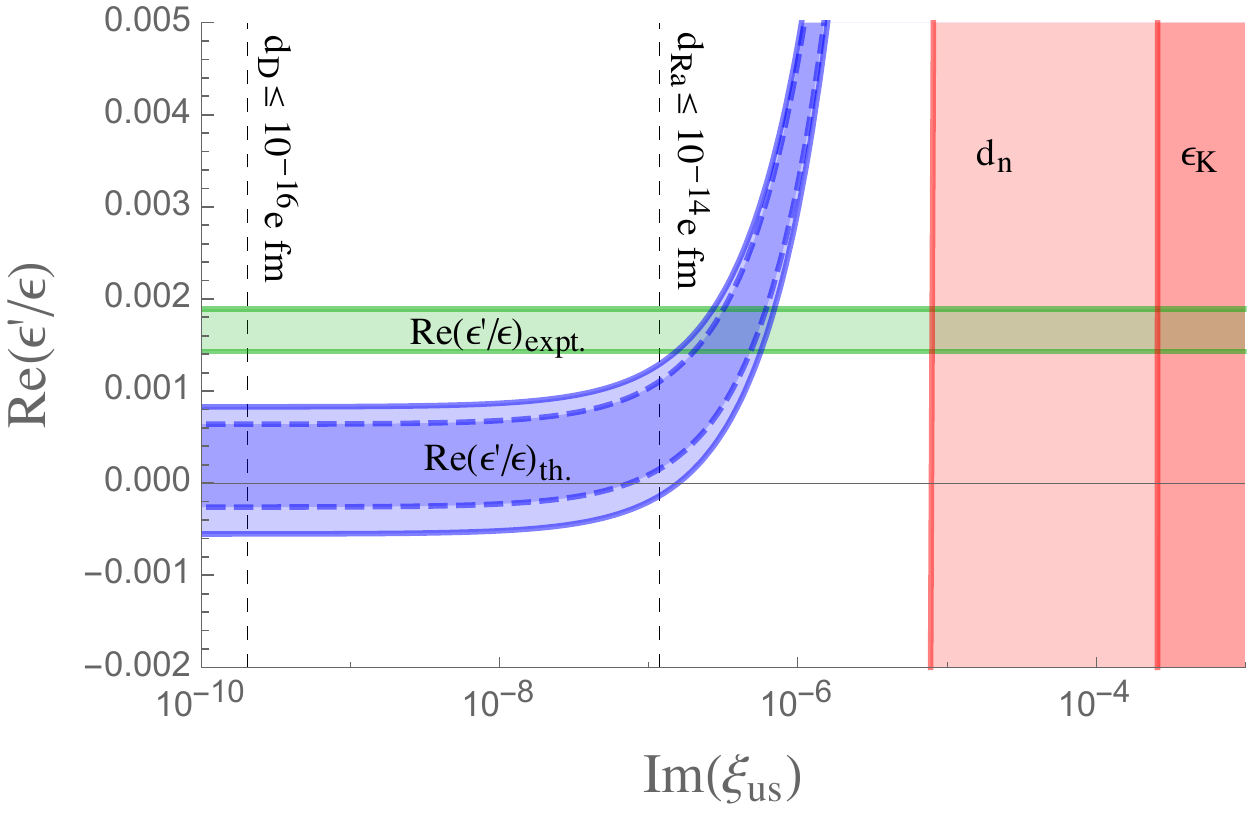}
\vspace{1.0cm}
\includegraphics[width=16cm]{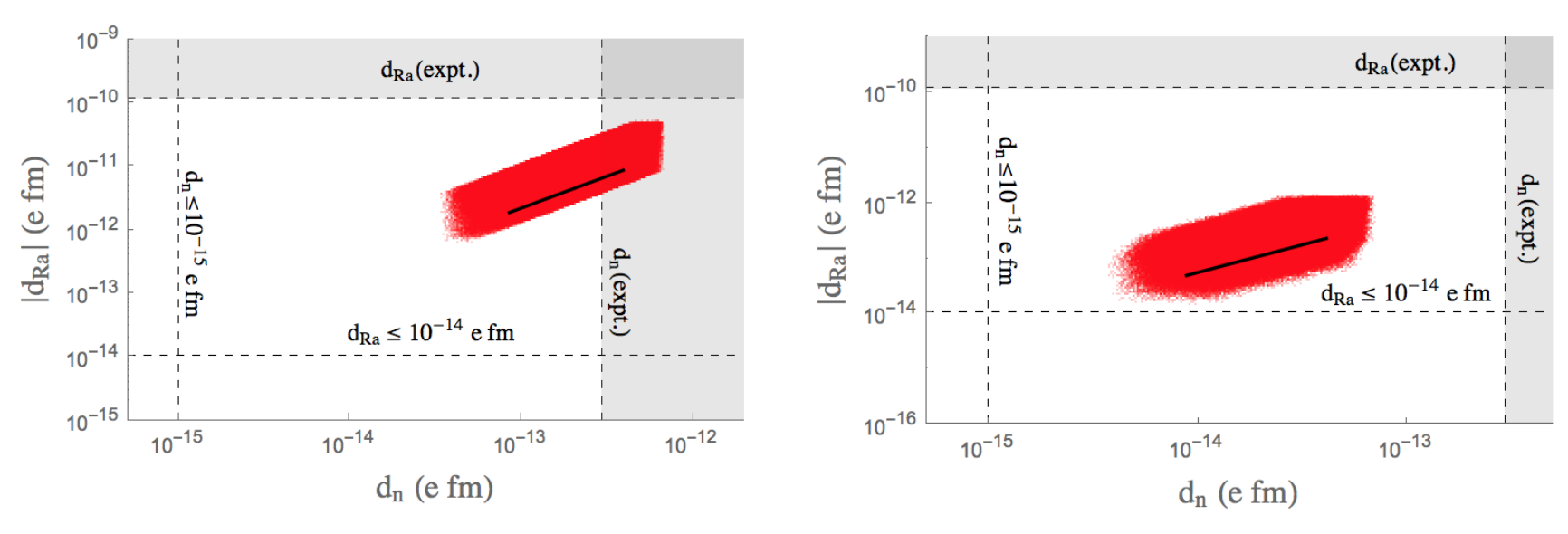}
\label{Fig1}
\vspace{-2cm}
\caption{\small The top-left (-right) panel shows the value of ${\rm Re}\,\ep'/\ep$ as a function of Im$\,\xi_{ud}$ (Im$\,\xi_{us}$). 
The solid and dashed blue bands indicate the theory prediction  (see text), while the experimental value is shown in green (all at $1\, \sigma$). The vertical lines indicate the current/future sensitivities of $\epsilon_K$ and  $d_{n,D,Ra}$ experiments, derived using the R-fit procedure.
The bottom-left (-right) panel shows the sizes of $d_{\rm Ra}$ and $d_n$, assuming a value for Im$\,\xi_{ud}$ (Im$\,\xi_{us}$) that solves the $\ep'/\ep$ discrepancy. The red points are generated by taking random values of the nuclear and hadronic matrix elements within their allowed ranges. The black lines result from taking the central values of these matrix elements.}\label{Fig:plots}
\vspace{-.0cm}
\end{figure}

Using the expressions in section \ref{sec:eps} we show $\ep'$ as a function of $\xi_{ud}$ and $\xi_{us}$ in the upper-left and -right panels of Fig. \ref{Fig:plots}, respectively. Here the green band indicates the experimental value, while the solid and dashed blue lines are theory predictions using the SM values of Ref.\ \cite{Bai:2015nea} and \cite{Buras:2015yba}, respectively.
These panels show that the tension can be alleviated when the couplings have sizes of $\Or(10^{-7}-10^{-6})$. Coefficients of this size  naively point towards a scale of $\Lambda=\Or(100\, {\rm TeV})$, although, in specific models this scale can be lowered by small model parameters. 
The same panels also show the constraints from $\ep_K$ and $d_n$, and future $d_{D,{\rm Ra}}$ sensitivities~\cite{Chupp:2014gka,Eversmann:2015jnk}.
 In principle, the stringent experimental limit on the mercury EDM also leads to strong constraints if one neglects theoretical uncertainties. 
However, here we follow the R-fit procedure~\cite{Charles:2004jd} to obtain limits. In this case the large nuclear and hadronic uncertainties allow for  cancellations which result in a vanishing $d_{\rm Hg}$ EDM.

As can be seen from the Figure, the current $\ep_K$ and $d_n$ limits do not rule out the region of interest. For both $\xi_{ud}$ and $\xi_{us}$ a future neutron, deuteron, or radium measurement would be able to probe the regions in which the $\ep'$ tension is resolved.  
To illustrate this more clearly, we show the allowed values of the  neutron and radium EDMs in the lower-left and -right panels of Fig.\ \ref{Fig:plots}. Here the black lines indicate the  neutron and radium EDMs in the case that $\xi_{ud,us}$ have the right size to solve the $\ep'$ discrepancy. The red points show the values once one accounts for hadronic and nuclear uncertainties.  For both $\xi_{ud}$ and $\xi_{us}$, one can see that the projected experimental sensitivities for  $d_n$ and $d_{\rm Ra}$ would allow one to rule out the entire parameter space. 
In summary, a right-handed explanation of the tension in $\ep'$ points to a scale of new physics of the order of $\Or (100\, {\rm TeV})$, which is not within reach of direct searches. Nonetheless, future EDM experiments would be able to falsify this scenario.

\section*{Acknowledgments}

I would like to thank the organizers of the $52^{\rm nd}$ Rencontres de Moriond for an interesting and enjoyable meeting. I am grateful to Vincenzo Cirigliano, Emmanuele Mereghetti, and Jordy de Vries for the collaboration on this work. This work was supported by the Dutch Organization for Scientific Research (NWO) through a RUBICON grant.

\section*{References}

\bibliography{bibliography}

\begin{thebibliography}{10}

\bibitem{AlaviHarati:2002ye}
A.~Alavi-Harati et~al.
\newblock {Measurements of direct CP violation, CPT symmetry, and other
  parameters in the neutral kaon system}.
\newblock {\em Phys. Rev. D}, 67:012005, 2003.
\newblock [Erratum: Phys. Rev.D70,079904(2004)].

\bibitem{Abouzaid:2010ny}
E.~Abouzaid et~al.
\newblock {Precise Measurements of Direct CP Violation, CPT Symmetry, and Other
  Parameters in the Neutral Kaon System}.
\newblock {\em Phys. Rev. D}, 83:092001, 2011.

\bibitem{Batley:2002gn}
J.~R. Batley et~al.
\newblock {A Precision measurement of direct CP violation in the decay of
  neutral kaons into two pions}.
\newblock {\em Phys. Lett. B}, 544:97--112, 2002.

\bibitem{Bai:2015nea}
Z.~Bai et~al.
\newblock {Standard Model Prediction for Direct CP Violation in K→ππ
  Decay}.
\newblock {\em Phys. Rev. Lett.}, 115(21):212001, 2015.

\bibitem{Buras:2015xba}
Andrzej~J. Buras and Jean-Marc G\'erard.
\newblock {Upper bounds on $\epsilon'/\epsilon$ parameters B$_{6}^{(1/2)}$ and
  B$_{8}^{(3/2)}$ from large N QCD and other news}.
\newblock {\em JHEP}, 12:008, 2015.

\bibitem{Buras:2016fys}
Andrzej~J. Buras and Jean-Marc G\'erard.
\newblock {Final state interactions in $K\rightarrow \pi \pi $ decays: $\Delta
  I=1/2$ rule vs. $\varepsilon '/\varepsilon $}.
\newblock {\em Eur. Phys. J.}, C77(1):10, 2017.

\bibitem{Kitahara:2016nld}
Teppei Kitahara, Ulrich Nierste, and Paul Tremper.
\newblock {Singularity-free next-to-leading order $\Delta$S = 1 renormalization
  group evolution and $\epsilon_K'/\epsilon_K$ in the Standard Model and
  beyond}.
\newblock {\em JHEP}, 12:078, 2016.

\bibitem{Bertolini:2000dy}
Stefano Bertolini, Jan~O. Eeg, and Marco Fabbrichesi.
\newblock {An Updated analysis of epsilon-prime / epsilon in the standard model
  with hadronic matrix elements from the chiral quark model}.
\newblock {\em Phys. Rev. D}, 63:056009, 2001.

\bibitem{Bijnens:2000im}
Johan Bijnens and Joaquim Prades.
\newblock {Epsilon-prime K / K epsilon in the chiral limit}.
\newblock {\em JHEP}, 06:035, 2000.

\bibitem{Pallante:2001he}
E.~Pallante, A.~Pich, and I.~Scimemi.
\newblock {The Standard model prediction for epsilon-prime / epsilon}.
\newblock {\em Nucl. Phys. B}, 617:441--474, 2001.

\bibitem{Bobeth:2016llm}
Christoph Bobeth, Andrzej~J. Buras, Alejandro Celis, and Martin Jung.
\newblock {Patterns of Flavour Violation in Models with Vector-Like Quarks}.
\newblock {\em JHEP}, 04:079, 2017.

\bibitem{Buras:2015kwd}
Andrzej~J. Buras and Fulvia De~Fazio.
\newblock {$\varepsilon'/\varepsilon$ in 331 Models}.
\newblock {\em JHEP}, 03:010, 2016.

\bibitem{Buras:2015jaq}
Andrzej~J. Buras.
\newblock {New physics patterns in $\varepsilon^\prime/\varepsilon$ and
  $\varepsilon_K$ with implications for rare kaon decays and $\Delta M_K$}.
\newblock {\em JHEP}, 04:071, 2016.

\bibitem{Kitahara:2016otd}
Teppei Kitahara, Ulrich Nierste, and Paul Tremper.
\newblock {Supersymmetric Explanation of CP Violation in $K\to \pi\pi$ Decays}.
\newblock {\em Phys. Rev. Lett.}, 117(9):091802, 2016.

\bibitem{Endo:2016aws}
Motoi Endo, Satoshi Mishima, Daiki Ueda, and Kei Yamamoto.
\newblock {Chargino contributions in light of recent $\epsilon'/\epsilon$}.
\newblock {\em Phys. Lett. B}, 762:493--497, 2016.

\bibitem{Crivellin:2017gks}
Andreas Crivellin, Giancarlo D'Ambrosio, Teppei Kitahara, and Ulrich Nierste.
\newblock {$K\to \pi \nu\overline{\nu}$ in the MSSM in light of the
  $\epsilon^{\prime}_K/\epsilon_K$ anomaly}.
\newblock {\em Phys. Rev. D}, 96(1):015023, 2017.

\bibitem{Cirigliano:2016yhc}
V.~Cirigliano, W.~Dekens, J.~de~Vries, and E.~Mereghetti.
\newblock {An $\epsilon'$ improvement from right-handed currents}.
\newblock {\em Phys. Lett. B}, 767:1--9, 2017.

\bibitem{Mohapatra:1974hk}
Rabindra~N. Mohapatra and Jogesh~C. Pati.
\newblock {Left-Right Gauge Symmetry and an Isoconjugate Model of CP
  Violation}.
\newblock {\em Phys. Rev. D}, 11:566--571, 1975.

\bibitem{Senjanovic:1975rk}
G.~Senjanovic and Rabindra~N. Mohapatra.
\newblock {Exact Left-Right Symmetry and Spontaneous Violation of Parity}.
\newblock {\em Phys. Rev. D}, 12:1502, 1975.

\bibitem{Alioli:2017ces}
S.~Alioli, V.~Cirigliano, W.~Dekens, J.~de~Vries, and E.~Mereghetti.
\newblock {Right-handed charged currents in the era of the Large Hadron
  Collider}.
\newblock {\em JHEP}, 05:086, 2017.

\bibitem{Peccei:1977hh}
R.~D. Peccei and Helen~R. Quinn.
\newblock {CP Conservation in the Presence of Instantons}.
\newblock {\em Phys. Rev. Lett.}, 38:1440--1443, 1977.

\bibitem{Bijnens:1983ye}
Johan Bijnens and Mark~B. Wise.
\newblock {Electromagnetic Contribution to Epsilon-prime/Epsilon}.
\newblock {\em Phys. Lett. B}, 137:245--250, 1984.

\bibitem{Buchalla:1995vs}
Gerhard Buchalla, Andrzej~J. Buras, and Markus~E. Lautenbacher.
\newblock {Weak decays beyond leading logarithms}.
\newblock {\em Rev. Mod. Phys.}, 68:1125--1144, 1996.

\bibitem{Blum:2012uk}
T.~Blum et~al.
\newblock {Lattice determination of the $K \to (\pi\pi)_{I=2}$ Decay Amplitude
  $A_2$}.
\newblock {\em Phys. Rev. D}, 86:074513, 2012.

\bibitem{Buras:2015yba}
Andrzej~J. Buras, Martin Gorbahn, Sebastian J\"{a}ger, and Matthias Jamin.
\newblock {Improved anatomy of ε′/ε in the Standard Model}.
\newblock {\em JHEP}, 11:202, 2015.

\bibitem{Cirigliano:2011ny}
Vincenzo Cirigliano, Gerhard Ecker, Helmut Neufeld, Antonio Pich, and Jorge
  Portoles.
\newblock {Kaon Decays in the Standard Model}.
\newblock {\em Rev. Mod. Phys.}, 84:399, 2012.

\bibitem{deVries:2012ab}
J.~de~Vries, E.~Mereghetti, R.~G.~E. Timmermans, and U.~van Kolck.
\newblock {The Effective Chiral Lagrangian From Dimension-Six Parity and
  Time-Reversal Violation}.
\newblock {\em Annals Phys.}, 338:50--96, 2013.

\bibitem{Bsaisou:2014oka}
J.~Bsaisou, Ulf-G. Mei{\ss}ner, A.~Nogga, and A.~Wirzba.
\newblock {P- and T-Violating Lagrangians in Chiral Effective Field Theory and
  Nuclear Electric Dipole Moments}.
\newblock {\em Annals Phys.}, 359:317--370, 2015.

\bibitem{Aoki:2016frl}
S.~Aoki et~al.
\newblock {Review of lattice results concerning low-energy particle physics}.
\newblock 2016.

\bibitem{Borsanyi:2013lga}
Sz. Borsanyi, S.~D�rr, Z.~Fodor, J.~Frison, C.~Hoelbling, et~al.
\newblock {Isospin splittings in the light baryon octet from lattice QCD and
  QED}.
\newblock {\em Phys.Rev.Lett.}, 111:252001, 2013.

\bibitem{Borsanyi:2014jba}
Sz. Borsanyi et~al.
\newblock {Ab initio calculation of the neutron-proton mass difference}.
\newblock {\em Science}, 347:1452--1455, 2015.

\bibitem{Brantley:2016our}
David~A. Brantley, Balint Joo, Ekaterina~V. Mastropas, Emanuele Mereghetti,
  Henry Monge-Camacho, Brian~C. Tiburzi, and Andre Walker-Loud.
\newblock {Strong isospin violation and chiral logarithms in the baryon
  spectrum}.
\newblock 2016.

\bibitem{Mereghetti:2010kp}
E.~Mereghetti, J.~de~Vries, W.~H. Hockings, C.~M. Maekawa, and U.~van Kolck.
\newblock {The Electric Dipole Form Factor of the Nucleon in Chiral
  Perturbation Theory to Sub-leading Order}.
\newblock {\em Phys. Lett. B}, 696:97--102, 2011.

\bibitem{Seng:2014pba}
Chien-Yeah Seng, Jordy de~Vries, Emanuele Mereghetti, Hiren~H. Patel, and
  Michael Ramsey-Musolf.
\newblock {Nucleon electric dipole moments and the isovector parity- and
  time-reversal-odd pion–nucleon coupling}.
\newblock {\em Phys. Lett. B}, B736:147--153, 2014.

\bibitem{deJesus:2005nb}
J.~H. de~Jesus and J.~Engel.
\newblock {Time-reversal-violating Schiff moment of Hg-199}.
\newblock {\em Phys. Rev.}, C72:045503, 2005.

\bibitem{Bsaisou:2014zwa}
J.~Bsaisou, J.~de~Vries, C.~Hanhart, S.~Liebig, Ulf-G. Mei{\ss}ner, D.~Minossi,
  A.~Nogga, and A.~Wirzba.
\newblock {Nuclear Electric Dipole Moments in Chiral Effective Field Theory}.
\newblock {\em JHEP}, 03:104, 2015.
\newblock [Erratum: JHEP05,083(2015)].

\bibitem{Ban:2010ea}
Shufang Ban, Jacek Dobaczewski, Jonathan Engel, and A.~Shukla.
\newblock {Fully self-consistent calculations of nuclear Schiff moments}.
\newblock {\em Phys. Rev. C}, 82:015501, 2010.

\bibitem{Dzuba:2009kn}
V.~A. Dzuba, V.~V. Flambaum, and S.~G. Porsev.
\newblock {Calculation of P,T-odd electric dipole moments for diamagnetic atoms
  Xe-129, Yb-171, Hg-199, Rn-211, and Ra-225}.
\newblock {\em Phys. Rev. A}, 80:032120, 2009.

\bibitem{Engel:2013lsa}
Jonathan Engel, Michael~J. Ramsey-Musolf, and U.~van Kolck.
\newblock {Electric Dipole Moments of Nucleons, Nuclei, and Atoms: The Standard
  Model and Beyond}.
\newblock {\em Prog. Part. Nucl. Phys.}, 71:21--74, 2013.

\bibitem{deVries2011b}
J.~de~Vries, R.~Higa, C.-P. Liu, E.~Mereghetti, I.~Stetcu, et~al.
\newblock {Electric Dipole Moments of Light Nuclei From Chiral Effective Field
  Theory}.
\newblock {\em Phys. Rev. C}, 84:065501, 2011.

\bibitem{Singh:2015aba}
Yashpal Singh and B.~K. Sahoo.
\newblock {Electric dipole moment of $^{225}$Ra due to P- and T-violating weak
  interactions}.
\newblock {\em Phys. Rev. A}, 92:022502, 2015.

\bibitem{Yamanaka:2017mef}
N.~Yamanaka, B.~K. Sahoo, N.~Yoshinaga, T.~Sato, K.~Asahi, and B.~P. Das.
\newblock {Probing exotic phenomena at the interface of nuclear and particle
  physics with the electric dipole moments of diamagnetic atoms: A unique
  window to hadronic and semi-leptonic CP violation}.
\newblock {\em Eur. Phys. J. A}, 53:54, 2017.

\bibitem{Afach:2015sja}
J.M. Pendlebury et~al.
\newblock {Revised experimental upper limit on the electric dipole moment of
  the neutron}.
\newblock {\em Phys. Rev. D}, 92(9):092003, 2015.

\bibitem{Graner:2016ses}
B.~Graner, Y.~Chen, E.~G. Lindahl, and B.~R. Heckel.
\newblock {Reduced Limit on the Permanent Electric Dipole Moment of Hg199}.
\newblock {\em Phys. Rev. Lett.}, 116(16):161601, 2016.

\bibitem{Bishof:2016uqx}
Michael Bishof et~al.
\newblock {Improved limit on the $^{225}$Ra electric dipole moment}.
\newblock {\em Phys. Rev. C}, 94(2):025501, 2016.

\bibitem{Parker:2015yka}
R.H. Parker et~al.
\newblock {First Measurement of the Atomic Electric Dipole Moment of
  $^{225}$Ra}.
\newblock {\em Phys. Rev. Lett.}, 114(23):233002, 2015.

\bibitem{Chupp:2014gka}
Timothy Chupp and Michael Ramsey-Musolf.
\newblock {Electric Dipole Moments: A Global Analysis}.
\newblock {\em Phys. Rev. C}, 91(3):035502, 2015.

\bibitem{Eversmann:2015jnk}
D.~Eversmann et~al.
\newblock {New method for a continuous determination of the spin tune in
  storage rings and implications for precision experiments}.
\newblock {\em Phys. Rev. Lett.}, 115(9):094801, 2015.

\bibitem{Charles:2004jd}
J.~Charles, Andreas Hocker, H.~Lacker, S.~Laplace, F.~R. Le~Diberder,
  J.~Malcl\`es, J.~Ocariz, M.~Pivk, and L.~Roos.
\newblock {CP violation and the CKM matrix: Assessing the impact of the
  asymmetric $B$ factories}.
\newblock {\em Eur. Phys. J. C}, 41:1--131, 2005.

\end{thebibliography}
\end{document}